\documentclass[multphys,vecphys]{svmult}

\usepackage{makeidx}         % allows index generation
\usepackage{graphicx}        % standard LaTeX graphics tool
\usepackage{multicol}        % used for the two-column index
\usepackage[bottom]{footmisc}% places footnotes at page bottom
\makeindex             % used for the subject index
\begin{document}

\title*{Tidal disruption and the tale of three clusters}
% Use \titlerunning{Tidal disruption of globular clusters} for an 
% abbreviated version of
% your contribution title if the original one is too long

\author{Guido De Marchi\inst{1} \and Luigi Pulone\inst{2} \and
Francesco Paresce\inst{3}}

\institute{ESA, Space Science Department, Noordwijk, Netherlands
\texttt{gdemarchi@esa.int} 
\and INAF, Observatory of Rome, Monte Porzio, Italy
\texttt{pulone@mporzio.astro.it} 
\and INAF, Rome, Italy \texttt{fparesce@inaf.it} }  

\maketitle

\section*{Abstract\footnote{Space limitations do not allow us
to present our work in full detail. We, therefore, include here a brief
summary of the main results. For the complete version of this paper,
the reader should refer to De Marchi et al. (2006) \cite{DPP06}.}} 
\label{sec:1}

How well can we tell whether a cluster will survive the Galaxy's tidal
forces? This is conceptually easy to do if we know the cluster's total
mass, mass structure and space motion parameters. This information is
used in models that predict the probability of disruption due to tidal
stripping, disc and bulge shocking \cite{GO97,DGA99,BM03}. But just how
accurate is the information that goes into these models and, therefore,
how reliable are their predictions? To understand the virtues and
weaknesses of these models, we have studied in detail three clusters
(NGC 6397, NGC 6712, NGC 6218) whose predicted interaction with the
galaxy is very different. We have used deep HST and VLT data to measure
the luminosity function (LF) of stars throughout the clusters in order
to derive a solid global mass function (GMF). The latter is the best
tell-tale of the strength and extent of tidal stripping operated by the
Galaxy. This is because the evaporation of stars from the cluster
causes a progressive flattening of the IMF \cite{VH97} and this effect
is enhanced by tidal stripping and disc/bulge shocking. Therefore, at
any time, the shape of the GMF must reflect the past interaction of the
cluster with the Galaxy. Since the three clusters that we have studied
have widely different probabilities of disruption (see the predicted 
times to disruption $T_\mathrm{d}$ in Table\,1), we expected to find
widely different GMFs. We indeed found that the GMF of the three
clusters is different, but not in the way predicted by the models, as
we explain below. 

To derive a reliable GMF, it is necessary to measure the stellar LF at
various locations in the cluster, in order to build a solid model of
the cluster's dynamics. Our accurate measurements in several bands
\cite{DPP00,ADFPPB01,DPP06} have allowed us to determine the GMF of the
three clusters for stars down to $\sim 0.2$\,M$_\odot$, where it is most
sensitive to tidal stripping. While the GMF of NGC 6397 is that typical
of globular clusters, with a peak at $\sim 0.3$\,M$_\odot$, NGC 6218
and NGC 6712 have very flat GMF. Near the half-mass radius, the LF that
we observed for these two clusters is severely depleted at the low-mass
end. But while the orbit of NGC 6712 is compatible with ``ferocious''
stripping, that of NGC 6218 is not \cite{DGCDOT96}. However, our
analysis shows that the orbit of NGC 6218 used in these models is not
realistic. In fact, more recent work based on the Hipparcos reference
system \cite{OBGT97} indicates for NGC 6218 an irregular orbit with
shorter perigalactic distance, more prone to extensive stripping. We
conclude that existing models of cluster disruption are limited by the
lack of information on the precise cluster space motions. However, they
benefit from an accurate knowledge of the GMF. Before the advent of
GAIA, which will make it possible to measure reliably the orbits of
many clusters, the only way forward to understand the evolution of the
globular cluster system rests on the accurate measurement of the GMF of
a large number of these objects. A telescope such as the VLT is
perfectly suited for this task.

\vspace*{-0.2 cm}
\begin{table}
\centering
\caption{Predicted times to disruption $T_{\rm d}$ in Gyr}
\label{tab:1}       % Give a unique label
\begin{tabular}{lccc}
\hline\noalign{\smallskip}
Reference & NGC\,6397\,\, & NGC\,6218\,\, & NGC\,6712\,\,  \\
\noalign{\smallskip}\hline\noalign{\smallskip}
Gnedin \& Ostriker 1997    & 2.1  & 14.5 & 0.3  \\
Dinescu et al. 1999        & 3.9  & 29.4 & 3.7  \\
Baumgardt \& Makino 2003   & 11.3 & 16.3 & 9.0  \\
\noalign{\smallskip}\hline
\end{tabular}
\end{table}

\vspace*{-0.5 cm}
\begin{figure}
\begin{minipage}{4.8cm}
\includegraphics[height=4.8cm]{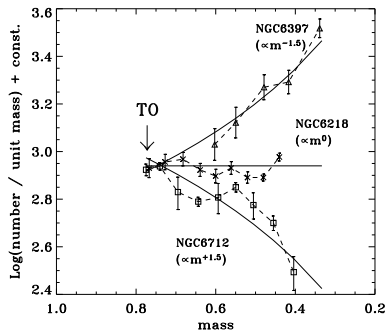} 
\end{minipage}
\hfill
\begin{minipage}{6cm}
\vspace*{-0.2cm} 
\caption{The solid lines show the best fitting power-law GMF as derived
using our dynamical models applied to a set of radial LFs. The symbols
joined by the dashed lines show the local  MF as measured near the
half-mass radius of each cluster, proving that the latter is a good
approximation to the GMF. While the time to disruption ($T_{\rm d}$) of
NGC 6712 justifies  its falling GMF, NGC 6218 should have a GMF as
steep as that of NGC 6397, given its very long $T_{\rm d}$.}
\label{fig:1}  
\end{minipage}
\end{figure}

%%%%%%%%%%%%%%%%%%%%%%%%%%%%%%%%%%%%%%%%%%%%%%%%%%%%%%%%%%%%%%%%%%%%%%  }

%%%%%%%%%%%%%%%%%%%%%%%%%%%%%%%%%%%%%%%%%%%%%%%%%%%%%%%%%%%%%%%%%%%%%%

\printindex

\vspace*{-1 cm}
\begin{thebibliography}{99.}
 

\bibitem{ADFPPB01} G. Andreuzzi, G. De Marchi, F. Ferraro, F. Paresce,
L. Pulone, R. Buonanno: A\&A, \textbf{372}, 851 (2001)

\bibitem{BM03} H. Baumgardt, J. Makino: MNRAS, \textbf{340}, 227 (2003)

\bibitem{DGCDOT96} B. Dauphole, M. Geffret, J. Colin, C. Ducourant, M. 
Odenkirchen, H. Tucholke: A\&A, \textbf{313}, 119 (1996)

\bibitem{DPP00} G. De Marchi, F. Paresce, L. Pulone: ApJ, \textbf{530}, 342
(2000)

\bibitem{DPP06} G. De Marchi, L. Pulone, F. Paresce: A\&A, \textbf{449}, 161
(2006)

\bibitem{DGA99} D. Dinescu, T. Girard, W. van Altena: AJ, \textbf{117},
1792 (1999)

\bibitem{GO97} O. Gnedin, J. Ostriker: ApJ, \textbf{486}, 581 (1997)

\bibitem{OBGT97} M. Odenkirchen, P. Brosche, M. Geffert, H. Tucholke:
  NewA, \textbf{2}, 477 (1997)

\bibitem{VH97} E. Vesperini, D. Heggie: MNRAS, 289, \textbf{898} (1997)


\end{thebibliography}
\end{document}